\documentclass[showpacs,prl,aps,twocolumn,floatfix]{revtex4}

\usepackage{bm}
\usepackage{amsmath}
\usepackage{amssymb}
\usepackage{latexsym} 
\usepackage{amsfonts} 
\usepackage{epsfig}
\usepackage{psfrag}

\newcommand{\bea}{\begin{eqnarray}}
\newcommand{\ea}{\end{eqnarray}}
\newcommand{\eea}{\end{eqnarray}}
\newcommand{\ord}{{\cal O}}

\begin{document}
  
\title{Quantum toy model for black-hole back-reaction} 

\author{Clovis Maia$^{1,2}$ and Ralf Sch\"utzhold$^{1,*}$}    

\affiliation{$^1$Institut f\"ur Theoretische Physik, 
Technische Universit\"at Dresden, 01062 Dresden, Germany}  

\affiliation{$^2$Instituto de F\'isica Te\'orica, 
Universidade Estadual Paulista, 01405-900, S\~ao Paulo, SP, Brazil} 

\begin{abstract}
We propose a simple quantum field theoretical toy model for black hole
evaporation and study the back-reaction of Hawking radiation onto
the classical background.
It turns out that the horizon is also ``pushed back'' in this situation 
(i.e., the interior region shrinks) but this back-reaction is not caused 
by energy conservation but by momentum balance. 
The effective heat capacity and the induced entropy variation can have 
both signs -- depending on the parameters of the model.
\end{abstract}


\maketitle

{\em Introduction}\quad 
%
Black holes are arguably the most simple and at the same time most 
intriguing objects in the universe. 
The no-hair theorem states that they can fully be described by a 
small set of parameters such as their mass $M$ and angular momentum $J$. 
Yet our standard picture of black holes contains many striking 
properties: 
Even though black holes should be completely black classically, 
they emit Hawking radiation due to quantum effects \cite{hawking}.
This evaporation process causes the black hole (horizon) to shrink 
(in the absence of infalling matter) due to the back-reaction of 
Hawking radiation.
Therefore, black holes possess a negative heat capacity \cite{charge}, 
i.e., the temperature grows with decreasing energy.  
Extrapolating this picture till the final stages of the evaporation, 
the black hole should end up in an explosion, where its temperature 
blows up and thus effects of quantum gravity should become important.   
Perhaps most fascinating is the observation that the second law of 
thermodynamics apparently \cite{thermo} requires to assign an entropy
$S$ to the black hole, which is determined by the horizon surface
area $A$ via $S=A/4$ (in natural units $\hbar=G=c=1$).

Taking the analogy between black holes and thermodynamics seriously 
provides a very consistent picture, which has been confirmed by various
gedanken experiments \cite{thermo,box} considering the construction of
heat engines with black holes etc. 
It almost seems as if nature was trying to give us some hints regarding 
the underlying structure which unifies quantum theory and gravity -- 
which we do not fully understand yet. 
In order to understand these hints better, it might be useful to ask 
the question of whether (and how) the aforementioned properties depend 
on the detailed structure of the Einstein equations or whether they are
more universal. 
For example, the study of condensed-matter based black hole analogues 
\cite{unruh,visser} shows that Hawking radiation is a fairly robust
quantum phenomenom \cite{universality}, which just requires the
occurrence of an effective horizon and is quite independent of the
Einstein equations.  
In contrast, the introduction of a black hole entropy with the desired
properties seems to rely on the Einstein equations. 

In the following, we try to further disentangle universal features from 
properties which are specific to black holes (e.g., Einstein equations, 
rotational symmetry, conserved ADM mass).
To this end, we propose a toy model which captures some of the relevant 
features of black holes and allows us to study the back-reaction of the 
emitted Hawking radiation onto the classical background solution. 

{\em Toy Model}\quad 
%
In the toy model we are going to discuss, the gravitational field will
be represented by a real scalar field $\psi$ in 1+1 dimensions with
the Lagrangian ($\hbar=1$) 
\bea
\label{L-psi}
{\cal L}_\psi
=
\frac12\left(\dot\psi^2-c_\psi^2[\partial_x\psi]^2\right)
-V(\psi)
\,.
\ea
With respect to the propagation speed $c_\psi$ of the $\psi$ field,
this form is Lorentz invariant.
The potential $V(\psi)$ is supposed to be very stiff, i.e., the field
$\psi$ is assumed to be heavy in the sense that it can be well
approximated by a classical field. 
For definiteness, we choose the sine-Gordon potential 
$V(\psi)\propto1-\cos(\psi/\psi_0)$, but other potentials admitting
stable solitonic solutions would also work.
The global ground state $\psi=0$ then corresponds to a vanishing
gravitational field whereas a kink (topological defect) models a black
(or white) hole horizon 
\bea
\label{kink}
\psi(x)=-4\psi_0\arctan\left(\exp\left\{-\xi[x-x_{\rm kink}]\right\}\right)
\,.
\ea
The position $x=x_{\rm kink}$ of the kink at rest is arbitrary and its width 
$1/\xi$ is determined by $V(\psi)$ and $c_\psi$. 
In comparison to other models of black holes 
(see, e.g., \cite{dilaton,balbinot}), the advantage of the above
set-up lies in the topologically protected stability and localization
of the kink, which behaves very similar to a particle
(see also \cite{domain}). 

In order to study Hawking radiation and its impact on the kink, we
consider a massless quantum field $\phi$ coupled to the heavy field
$\psi$ via the coupling constant $g$
\bea
\label{L-phi}
{\cal L}_\phi
=
\frac12\left([\partial_t\phi+g\psi\partial_x\phi]^2-
c_\phi^2[\partial_x\phi]^2\right)
\,.
\ea
Note that the velocity $c_\phi$ of the light (massless) field may
differ from $c_\psi$. 
The propagation of the light field $\phi$ in the approximately
classical background $\psi$ is completely analogous to that in a
gravitational field described by the 
Painlev{\'e}-Gullstrand-Lema{\^\i}tre metric 
(cf.~\cite{unruh,visser})
\bea
\label{PGL}
ds^2=\left(c_\phi^2-v^2\right)dt^2-2v\,dt\,dx-dx^2
\,,
\ea
where $v=g\psi$ denotes the local velocity of freely falling frames. 
A horizon occurs if this velocity $v$ exceeds the speed of light
$c_\phi$. 
Based on the analogy to gravity, we may also derive the pseudo
energy-momentum tensor of the $\phi$ field with respect to the above
metric $g^{\mu\nu}$ 
\bea
\label{pseudo}
T_{\mu\nu}
=
\frac{2}{\sqrt{-g}}\,
\frac{\delta{\cal A}_\phi}{\delta g^{\mu\nu}}
=
(\partial_\mu\phi)(\partial_\nu\phi)-
\frac12\,g_{\mu\nu}\,(\partial_\rho\phi)(\partial^\rho\phi)
\,.
\ea
The associated energy density $T_0^0$ of the light field 
\bea
\label{H-phi}
{\cal H}_\phi
=
\frac12\left([\partial_t\phi]^2+
\left(c_\phi^2-v^2\right)[\partial_x\phi]^2\right)
\ea
contains negative parts beyond the horizon $v^2>c_\phi^2$.
Of course, this is precisely the reason why effects like Hawking
radiation are possible \cite{birrell}.

However, an energy density which is not bounded from below seems
unphysical and typically indicates instabilities (already on the
classical level). 
In order to avoid this problem, we may add an extra term which does
not modify the linearized low-energy behavior of our model 
\bea
\label{L-reg}
{\cal L}_\phi^{\rm reg} 
=
{\cal L}_\phi-\alpha^2\left(c_\phi^2-v^2\right)^2
[\partial_x\phi]^4
-\frac{1}{16\alpha^2}
\,,
\ea
but generates a positive definite energy density  
\bea
\label{H-reg}
{\cal H}_\phi^{\rm reg} 
=
\frac12\left(
[\partial_t\phi]^2+
\left[
\alpha
(c_\phi^2-v^2)[\partial_x\phi]^2 
+\frac{1}{4\alpha}
\right]^2
\right)
\,.
\ea
In the exterior region $c_\phi^2>v^2$, the classical ground state is
still given by $\phi=0$, but beyond the horizon
$c_\phi^2<v^2$, we have
$2\alpha(\partial_x\phi)=(v^2-c_\phi^2)^{-1/2}$.  
Thus, the classical ground state profile would not be differentiable
at the horizon, i.e., the term $[\partial_x\phi]^2$ in the energy
density, for example, would be ill-defined. 
This problem can be cured by adding another term (which again does 
not modify the low-energy behavior) and we finally arrive at the 
total Lagrangian of our toy model 
\bea
\label{L-tot}
{\cal L}_{\rm full} 
=
{\cal L}_\psi
+
{\cal L}_\phi^{\rm reg} 
-\beta^2[\partial_x^2\phi]^2
\,.
\ea
The last term smoothens the classical ground state profile at the
horizon and induces a super-luminal dispersion relation 
$(\omega+vk)^2=c_\phi^2k^2+2\beta^2k^4$ at large
wavenumbers.

{\em Back-reaction}\quad 
%
The equation of motion of the light field can be derived from the
Lagrangian above 
\bea
\label{eom-phi}
(\partial_t+v\partial_x)(\partial_t+\partial_x v)\phi=
c_\phi^2\partial_x^2\phi
+\ord(\partial_x^4)
\,,
\ea
where $\ord(\partial_x^4)$ denote the higher-order $\alpha$ and
$\beta$ terms we added for stability and regularity reasons. 
Similarly, the heavy field evolves according to 
\bea
\label{eom-psi}
\ddot\psi-c_\psi^2\partial_x^2\psi
=V'(\psi)
-
g[\partial_t\phi+g\psi\partial_x\phi]\partial_x\phi
+\ord(\partial_x^4)
\,.
\ea
From the full set of equations, we see that the kink profile in
Eq.~(\ref{kink}) together with $\phi=0$ exactly solves the classical
equations of motion (though it is not the ground state). 
However, the impact of quantum fluctuations changes this picture:  
For $2\pi g\psi_0>c_\phi$, the kink acts as a black hole horizon and
thus emits Hawking radiation.
Of course, the energy/momentum given off must come from somewhere and
hence this quantum effects should have some impact on the classical
kink background. 

In order to estimate the quantum back-reaction, we quantize the fields 
$\phi\to\hat\phi$ as well as $\psi\to\hat\psi$ and employ a mean-field 
expansion $\hat\psi=\psi_{\rm cl}+\delta\hat\psi$ where 
$\psi_{\rm cl}$ denotes the classical kink profile in Eq.~(\ref{kink})
and $\delta\hat\psi$ as well as $\hat\phi$ are supposed to be small
(i.e., $\hat\phi,\delta\hat\psi\ll\psi_{\rm cl}$).
Taking the expectation value of Eq.~(\ref{eom-psi}) and comparing it 
with Eq.~(\ref{pseudo}), we find that the lowest-order contributions
of the quantum back-reaction force are just given by the expectation
value of the pseudo energy-momentum tensor \cite{back-psi}
\bea
\label{back}
\left[\partial_t^2-c_\psi^2\partial_x^2-V''(\psi_{\rm cl})\right]
\langle\delta\hat\psi\rangle
\approx
-g\langle\hat T^0_1\rangle
\,.
\ea
Remembering the covariant energy-momentum balance 
\bea
\label{pseudo-balance}
\nabla_\mu T^\mu_\nu=\frac{1}{\sqrt{-g}}\,\partial_\mu
\left(\sqrt{-g}\,T^\mu_\nu\right)-\frac12\,T^{\alpha\beta}\,
\partial_\nu g_{\alpha\beta}=0
\,,
\ea
we find that $\langle\hat T^0_1\rangle$ denotes the momentum density 
$\pi_\phi\phi'$, which varies with position in general. 
In contrast, the energy flux $\langle\hat T^1_0\rangle$ measured with
respect to the stationary frame is constant 
$\partial_x\langle\hat T^1_0\rangle=0$
for a kink at rest. 

Fortunately, the expectation value $\langle\hat T^\mu_\nu\rangle$ can 
be calculated analytically for a scalar field in 1+1 dimensions. 
In the Unruh state (which is the appropriate state for describing 
black-hole evaporation), one obtains \cite{trace}
\bea
\label{trace}
\langle\hat T^0_1\rangle
=
\frac{4vc_\phi(\kappa^2-[v']^2-\gamma v v'')-
\kappa^2(c_\phi+v)^2}{48\pi c_\phi^3\gamma^2}
\,,
\ea
with $\gamma=1-v^2/c_\phi^2$ and the effective surface gravity $\kappa$
determining the Hawking temperature 
\bea
\label{kappa}
T_{\rm Hawking}=\frac{\kappa}{2\pi}=
\frac{1}{2\pi}\left(\frac{dv}{dx}\right)_{v^2=c_\phi^2}
\,.
\ea
Note that $\langle\hat T^0_1\rangle$ calculated in the Unruh state  
is regular across black-hole horizon $v=-c_\phi$, but singular at
the white hole horizon $v=+c_\phi$. 
(The Israel-Hartle-Hawking state would be regular at both 
horizons.) 
Far away from the kink/horizon $v\to0$, we just get the usual thermal
flux $\langle\hat T^0_1\rangle=-\kappa^2/(48\pi c_\phi)$.

The corrections induced by the quantum back-reaction can be visualized
by incorporating them into an effective potential $V_{\rm eff}$ via  
\bea
\label{potential}
V_{\rm eff}'(\psi)=V'(\psi_{\rm cl})
-g\langle\hat T^0_1\rangle
\,. 
\ea
For the classical potential $V(\psi)$, all minima 
$\psi\in 2\pi\psi_0\mathbb Z$ occur at the same energy $V=0$. 
However, the effective potential $V_{\rm eff}$ is distorted such that
the central minimum is lower than the next one describing the black
hole interior $V_{\rm eff}(\psi=0)<V_{\rm eff}(-2\pi\psi_0)$. 
In this sense, the exterior region is effectively energetically
favorable and thus the horizon starts to move inwards, i.e., the black
hole shrinks. 
Alternatively, the same result can be derived directly from
Eq.~(\ref{back}) via classical time-dependent perturbation theory
around the kink solution.  
The differential operator on the left-hand side of Eq.~(\ref{back})
possesses a continuum of gapped propagating (delocalized) modes with 
$\omega^2>0$ and one localized zero-mode $\propto1/\cosh(\xi[x-x_{\rm kink}])$ 
with $\omega=0$, which just corresponds to a translation of the kink
position \cite{kink-modes}.  
After expanding the source term $-g\langle\hat T^0_1\rangle$ in
Eq.~(\ref{back}) into these modes, the perturbations in the continuous
spectrum $\omega^2>0$ just propagate away from the kink -- whereas the
spatial overlap between $-g\langle\hat T^0_1\rangle$ and the zero-mode
determines the acceleration $\ddot x_{\rm kink}<0$ of the kink position. 

{\em Energy and Momentum}\quad 
%
In contrast to the fluid analogues for black holes (with a steady
in- and out-flow of energy and momentum), for example, the
kink considered here represents a well localized object, which allows
us to ask the question of where the force pushing back the horizon
comes from. 
In general, the contribution of the $\phi$ field to the total
energy-momentum tensor $\partial_\mu{\cal T}^{\mu\nu}=0$ is different
from the pseudo energy-momentum tensor $\nabla_\mu T^{\mu\nu}=0$
defined with respect to the effective metric~(\ref{PGL}), which
complicates the analysis \cite{stone}. 
Fortunately, these difficulties are absent in our toy model where the
mixed components of both tensors coincide 
${\cal T}^\mu_\nu=T^\mu_\nu$. 
The energy density $T_0^0$ is given by Eq.~(\ref{H-phi}) and the
classical expression for the momentum flux density just reads  
$T_1^1=-T_0^0$ due to conformal invariance of the scalar field in 1+1
dimensions. 
Note, however, that the quantum expectation values differ due to the
trace anomaly \cite{trace}.
The energy flux density 
$T_0^1=\dot\phi\,\partial{\cal L}/\partial\phi'$ is given by 
$T_0^1=\dot\phi[v\dot\phi+(v^2-c_\phi^2)\phi']$ and differs from the 
momentum density $T_1^0$ in Eqs.~(\ref{eom-psi}) and (\ref{back})
for $v\neq0$.  

Far away from the kink, we may estimate the above quantities by
employing the geometric-optics approximation and replacing 
$\dot\phi\to\Omega$ and $\phi'\to k$.
For solutions of the dispersion relation 
$(\Omega+vk)^2=c_\phi^2k^2+\ord(k^4)$ corresponding to the outgoing 
Hawking radiation and its infalling partner particles, the energy
density per normalized amplitude $T_0^0=c\Omega^2/(c-|v|)$ changes its
sign at the horizon, cf.~Eq.~(\ref{H-phi}). 
The energy flux density $T_0^1=c\Omega^2$ is constant and positive 
everywhere (which is even true beyond the geometric-optics
approximation). 
Note that $\Omega$ is conserved as we are considering a
quasi-stationary scenario. 
Thus, the total energy budget is balanced since the outgoing Hawking
radiation carries away positive energy, but the infalling partners
have a negative energy. 

The momentum density $T_1^0=-c\Omega^2/(c-|v|)^2$, on the other hand,  
turns out to be negative everywhere -- or more precisely, far away
from the kink, cf.~the exact expression (\ref{trace}) with
$\Omega\sim\kappa$. 
Thus the momentum flux density $T_1^1=-c\Omega^2/(c-|v|)$, i.e., the
pressure, also changes sign at the horizon.
(The trace anomaly vanishes in the asymptotic region $v'=v''=0$ far
away from the kink where the geometric-optics approximation applies 
$T_1^1=-T_0^0$.) 
Consequently, while the Hawking particles carry away positive momentum
and push back the kink, their infalling partner particles act in the
opposite way and pull on the kink.
In summary, the momentum is not balanced and thus the kink starts to
move, i.e., the black-hole interior region shrinks.

{\em Thermodynamics}\quad 
%
The application of thermodynamic concepts to our toy model (in analogy 
to real black holes) presents some difficulties and ambiguities:
Considering the heat capacity $C=dE/dT$, for example, we would
associate $T$ with the Hawking temperature (\ref{kappa}).
The variation of the internal energy $dE$, however, could be identified
with the heat given off by the Hawking radiation 
$dE=\delta Q\propto\kappa^2dt$ or with the change of the kinetic
energy of the kink $E=M_{\rm eff}\dot x_{\rm kink}^2/2$ 
(for $\dot x_{\rm kink}^2\ll c^2_\psi$). 
Since the kink does not possess a conserved ADM mass, these quantities
will be different in general.  
Either way, the heat capacity $C=dE/dT$ could be positive as well as
negative since the Hawking temperature can be increased $dT>0$ or
decreased $dT<0$ by the quantum back-reaction of the evaporation
process. 
There are several different effects:
Due to the distortion of the effective potential $V_{\rm eff}$, the
shape of the kink deviates from the classical profile~(\ref{kink}). 
This deviation is governed by the aforementioned continuum modes
$\omega^2>0$. 
Furthermore, the kink starts to move -- which is described by the
zero-mode. 
The motion of the kink, in turn, implies a Doppler shift of the
Hawking radiation.
Finally, even in the rest frame of the kink, the position of the
horizon $x_h$ changes since the kink velocity $\dot x_{\rm kink}$ effectively
reduces the local frame-dragging speed $v$ and therefore the surface
gravity $\kappa=v'(x_h)$ may change. 
As a result of all these effects, the heat capacity depends on many
parameters ($c_\phi$, $c_\psi$, and $g\psi_0$ etc.) and may assume
negative as well as positive values.
In order to demonstrate this sign ambiguity, let us consider the case 
$c_\psi\gg c_\phi$ for simplicity. 
In this limit, the continuum modes $\omega^2>0$ are very fast and
hence the change of the shape of the kink can be neglected, i.e.,
the quantum back-reaction induces a rigid motion of the kink only. 
As another simplification, the transformation of the $\psi$-field into
the rest frame of the kink is just a Galilei transformation due to
$c_\psi\gg c_\phi$.
The new horizon position is then simply determined by  
$v(x_h)=-c_\phi+\dot x_{\rm kink}$. 
Linearizing this equality together with $\kappa=v'(x_h)$, 
we find that the variation $\delta\kappa$ of the surface gravity
induced by the acceleration of the kink $\delta\dot x_{\rm kink}$ is determined
by $\delta\kappa=v''(x_h)\delta\dot x_{\rm kink}/\kappa$. 
Since $v''(x_h)$ can be positive or negative (depending on the
relation between $c_\phi$ and $g\psi_0$), the temperature measured in
the rest frame of the kink could change in both directions. 
The temperature in the laboratory frame acquires an additional Doppler
shift, which is given by $\delta\kappa=-\kappa\delta\dot x_{\rm kink}/c_\phi$.
The relative strength of the two competing effects (Doppler shift and
horizon displacement) is given by $c_\phi v''/(v')^2$, which can be
above or below one. 
Ergo, both temperatures (in the kink frame and in the laboratory
frame) may increase or decrease due to the back-reaction of Hawking
radiation, i.e., the heat capacity can be positive or negative 
(or even infinite -- at the turning point where $\delta T=0$).  

Similar ambiguities apply to the entropy $dS=dE/T$. 
Choosing $dE=\delta Q\propto\kappa^2dt$ just reproduces the entropy
balance of the Hawking radiation in the exterior region -- which is of
course indeed thermal. 
Inserting the kinetic energy $E=M_{\rm eff}\dot x_{\rm kink}^2/2$, on the other
hand, we could violate the $2^{\rm nd}$ law since the kink can be
slowed down by incident coherent radiation (carrying zero entropy). 

{\em Conclusions}\quad 
%
Modeling the black hole (horizon) by a stable topological defect in
the form of a kink, we were able to derive the quantum back-reaction
of the resulting evaporation process. 
It turns out that the kink/horizon is also pushed inwards as in a
real black hole but, in contrast to the gravitational case, this 
back-reaction force is not caused by energy conservation but by
momentum balance. 
Energetically, the expansion of the horizon would be favorable because 
the minimum energy density in exterior region $\phi=\psi=0$ is far
above $1/(4\alpha)^2>0$ the ground state in the interior region.  
Hence, going beyond the linear analysis performed here, one might
suspect that the $\phi$ field approaches its ground state via
non-linear (quantum) instabilities until the evaporation stops. 

Further thermodynamical concepts such as heat capacity or entropy
(variation) cannot be defined unambiguously and can have both signs --
depending on the considered parameters \cite{fate}. 
Together with the results in \cite{balbinot}, our calculations and the
energy-momentum considerations above suggest that Hawking radiation and
the resulting back-reaction force ``pushing'' the horizon inwards may
be universal -- whereas the heat capacity and the entropy concept
strongly depend on the underlying structure (e.g., Einstein
equations). 
Note that in the Israel-Hartle-Hawking state with the expectation
value being 
$\langle\hat T^0_1\rangle=
v(\kappa^2-[v']^2-\gamma v v'')/(12\pi c_\phi^2\gamma^2)$, the horizon
is still pushed inwards -- i.e., it does not correspond to the thermal
equilibrium state for the combined system 
[kink in Eq.~(\ref{kink}) plus $\phi$ field].

R.~S.~acknowledges valuable discussions with Ted Jacobson, Bill Unruh,
Renaud Parentani, and others at the workshop {\em From Quantum to
  Emergent Gravity: Theory and Phenomenology} 
(SISSA, Trieste, Italy 2007) and support by the Emmy-Noether Programme
of the German Research Foundation (DFG, SCHU~1557/1-2). 
C.~M.~is indebted to G.~Matsas for the support, the ITP at TU Dresden
for the hospitality and Funda\c c\~ao de Amparo \`a Pesquisa do
Estado de S\~ao Paulo for financial support.

$^*$\,{\sf schuetz@theory.phy.tu-dresden.de}



\begin{thebibliography}{499}
\bibitem{charge}
For simplicity, we are considering black holes without charge $Q=0$ and 
angular momentum $J=0$.
%
Otherwise we would have to distinguish different heat capacities in 
analogy to $c_p$ and $c_V$ in thermodynamics, as can be seen by comparing 
the first law of thermodynamics $dE=TdS-pdV$ 
with the first law of black hole dynamics  
$dM=\kappa dA/(8\pi)+\Omega_H dJ+\Phi dQ$ 
where $\Omega_H$ is the angular velocity (at the horizon) 
and $\Phi$ is the (electrostatic) potential.
\bibitem{back-psi}
In a similar way, one could calculate the back-reaction of the quantum
fluctuations of the $\psi$ field itself. 
%
However, symmetry arguments show that these corrections do not
generate any force on the kink.
\end{thebibliography}
\end{document}